# Using local GND density to study SCC initiation

Naganand Saravanan[1], Phani S Karamched[1], Junliang Liu[1], Claire Rainasse[2], Fabio Scenini[3], Sergio Lozano-Perez[1]

[1]Department of Materials, University of Oxford, Parks Road, OX1 3PH, Oxford, UK
[2]EDF R&D, Avenue des Renardieres, Moret sur Loing Cedex, 77818, France
[3]The University of Manchester, Department of Materials, Materials Performance Centre, Manchester, M13 9PL, Manchester, UK

## Abstract

Strain and geometric necessary dislocations (GNDs) have been mapped with nm-resolution around grain boundaries affected by stress corrosion cracking (SCC) or intergranular oxidation with the aim of clarifying which local conditions that trigger SCC initiation of Alloy 600 in primary water reactor (PWR) water environment. Regions studied included the cracked and uncracked portion of the same SCC-affected grain boundaries and a comparable grain boundary in the as-received condition. High-resolution "on-axis" Transmission Kikuchi Diffraction (TKD) was used to generate strain and GND density based on the cross-correlation image processing method to probe shifts of specific zone axis in the TKD patterns from all regions. All cracked boundaries analyzed had local GND densities higher than $1 \times 10^{16} \, m^{-2}$. Similar grain boundaries, from as-received samples had GNDs of $5 \times 10^{14} \, m^{-2}$, while an intermediate level was found in the oxidized but uncracked portion of the same GB. Results, together with a discussion on the advantages and limitations of the approach, will be presented.

# 1. Introduction

Stress corrosion cracking (SCC) is a degradation mechanism that requires a specific environment and tensile stress, whether applied or residual, and this form of degradation has been reported in structural materials in nuclear reactor environment [1–4]. Microscopic cracks can develop due to the combination of primary water reactor (PWR) water environments, stresses on the material, and significantly limit the life span of the reactor components as a whole. Alloy 600 is a Nickel based alloy widely used as components in primary circuit of PWRs such as steam generator tubing and control rod driving mechanism (CRDM) nozzles due to their excellent corrosion resistance and good mechanical properties in high temperature water. However, this alloy is highly susceptible to Intergranular Stress Corrosion Cracking (IGSCC). There has been several research groups mainly focusing on IGSCC in Alloy 600 for over two decades now [5–8]. Nevertheless, there is still not a general understanding of the controlling SCC mechanisms and there is no valid general model for crack initiation or propagation.

In previous SCC propagation studies [9–12], various electron microscopy techniques such as scanning electron microscopy (SEM), transmission electron microscopy (TEM) [13,14], electron back scattered diffraction (EBSD) and off-axis transmission Kikuchi diffraction (TKD) [15] have been used to study oxide structure, chemistry, crack morphology, matrix and deformation around the crack tip, orientation of the grains around the crack etc. Kernel average misorientation (KAM) after EBSD mapping has been traditionally used to identify regions of high strain and correlate them with SCC initiation. Lattice rotations are measured locally and differences smaller than a degree can be detected. As an example, Ehrnsten et al. [16] used it to reveal the effect of cold-work on creating local high-strain areas, which is known to facilitate SCC initiation [17]. This method, however, remains qualitative, since it is highly dependent on the parameters used for the EBSD acquisition (*e.g.* step size) and cannot offer a lateral resolution better than the technique itself (>100 nm). Some researchers [2,18,19] used Digital Image Correlation (DIC) to correlate the microstructural scale plastic strain and misorientation. DIC, even when performed in an SEM, has a limited lateral resolution (*e.g.* 5 µm mesh in [6]), which conditions its strain measurements and might not highlight key regions of high strain localization. So, to summarize, differences in the acquisition parameters, often related to the step size, grain orientation chosen and the maximum resolution attainable, condition the use and limit the comparison between different works and materials. A way of overcoming the resolution problem is to perform on-axis TKD on TEM samples [20], so that KAM maps can be obtained with nm resolution. KAM maps can be used to measure matrix rotation gradients locally, which can be qualitatively related to dislocation densities [21]. However, although TKD KAM maps provide the necessary resolution to analyze the typical local deformation scales around grain boundaries and crack tips (tens of nm), it remains qualitative when it comes to comparison between different works, due to differences in step size, grain orientation, phase, *etc*. Geometrically necessary dislocations (GND) represent the excess dislocations stored within a Burger's circuit and contribute to lattice curvature, are accumulated in strain gradient fields caused by geometrical constraints of the crystal lattice [22], and can be calculated based on the shifts of specific zone axis in the TKD patterns [23]. Ruggles [21] studied that geometrically necessary dislocations (GNDs) associated with long range distortion gradient relating to heterogeneous deformation. Meisnar *et al.* [15] first used off-axis TKD to study localized deformation around SCC crack tips in stainless steel and Shen *et al.* [24] extended it to Alloy 600. They used average Misorientation (MO) maps to describe the size and extent of the localized deformation zone, relating it to local dislocation densities, although only qualitatively. It is, however, still not understood to what extent the knowledge of GNDs can contribute to better understanding SCC initiation, or whether they could highlight some form of density threshold that would allow stress localization to reach the minimum required to fracture an oxidized grain boundary [25,26]; for this reason a more in-depth study is necessary. In this paper, we consider the possibility of using GNDs and strain maps calculated from high-resolution TKD patters (details in next section) to characterize the build-up of dislocations around grain boundaries and its relation to intergranular crack initiation.

TKD uses electron transparent samples, prepared in the same way as a TEM sample, instead of polished bulk samples for EBSD. The Kikuchi diffraction patterns in TKD are obtained by electron beam transmitting through the sample and not by backscattering as in the case of EBSD, thus reducing the electron interaction volume and leading to improvement in lateral resolution. It has been reported that the lateral resolution is improved from 50 nm in the case of EBSD to 10 nm in TKD [27], which can be further enhanced by using "on-axis" detectors [28]. The other advantage of TKD is that it uses the same hardware as EBSD except that preparation of the sample and positioning in the chamber varies.

There is limited information on the relationship between strain and crystallographic features of different regions of the grain boundary. It is important to characterize the grain boundary because SCC is predominantly intergranular and investigate the local changes that make a grain boundary susceptible. This study includes strain measurements and GNDs measurements in three different regions of similar grain boundaries (a) the portion of a grain boundary that cracked (b) the portion of the same grain boundary that did not crack (c) a similar grain boundary in the as-received material and then cracked grain boundary from different region of the sample. Two cracked grain boundaries were characterized for improved statistics. These samples were chosen with the idea of obtaining the minimum amount (threshold value) of GNDs required for the crack to initiate, and attempt to use the measured elastic strain and stress to elucidate the mode of cracking.

## 2. Experimental

### 2.1 Material used and SCC test

The material used for this study was Alloy 600 whose composition is listed in Table 1. The material was solution annealed for 30 minutes at 1150°C followed by water quenching in order to limit the formation of intergranular carbides and therefore to increase the IGSCC susceptibility; however, complete suppression of carbide is never possible [29,30]. The average grain size of the material was measured to be 40±16 µm. The SCC test was conducted using a refreshed autoclave in simulated PWR water conditions with 2.7 ppm dissolved $H_2$ at 345°C for 800 hours. With these conditions the oxidizing potential of the samples was slightly in the reducing side of the Ni/NiO transition, thus corresponding to conditions where SCC initiation is most favourable [30,31]. The geometry of the samples used for this study was a bar 60 mm long and with a square cross section 10 mm × 7 mm that was actively stressed under 4-point bending at 310 MPa. The superficial stress on the sample was calculated using the ASTM G39-99 procedure. Throughout the test the stress was maintained constant by ensuring a constant load. Before testing the surface stressed in tension was ground with SiC paper, polished with 1 µm diamond suspension and finally polished using colloidal silica (0.3 µm). OPS was used to remove the ultrafine grained layer that is present during mechanical polishing and increase the SCC susceptibility of the sample [17,32], thus facilitating the SCC initiation studies. Post- autoclave test analyses were done on the sample revealed an average crack depth of 5 ± 1 µm from 10 different cracks analyzed.

**Table 1**

Chemical content of alloy used in this study (%wt)

| Ni | Cr | Fe | C | Mn | S | Si | Cu | Al | Ti | P |
|---|---|---|---|---|---|---|---|---|---|---|
| Bal. | 16.30 | 9.20 | 0.07 | 0.17 | <0.001 | 0.29 | 0.01 | 0.19 | 0.19 | 0.007 |

## 2.2 Methodology

### 2.2.1 Site-specific sample preparation using the focused ion beam (FIB)

Sample preparation is essential for the success of TKD analysis. A dual beam Zeiss Crossbeam 540 FIB-SEM was used for lifting-out the samples containing the regions of interest. A final low-energy cleaning (5 kV, 250 pA) was performed on all the samples in order to eliminate any damage on surface induce due to FIB milling. The quality of the TKD patterns and the maximum achieved spatial resolution both dependent on the thickness of the sample. The ideal thickness depends on the material composition and sample, being ~50 nm for the alloy studied and the details on the sample preparation procedure can be found on [33]. TEM samples were prepared from a grain boundary showing partial intergranular SCC along its length. In this way, two samples could be extracted: one containing a portion of cracked grain boundary and another with a portion of the same grain boundary but uncracked, as shown in Figure 1. Two different cracked grain boundaries were studied for increased statistical relevance. An extra sample was prepared from the as-received material (before autoclave testing) to characterize the dislocation densities around similar grain boundaries before deformation and cracking.

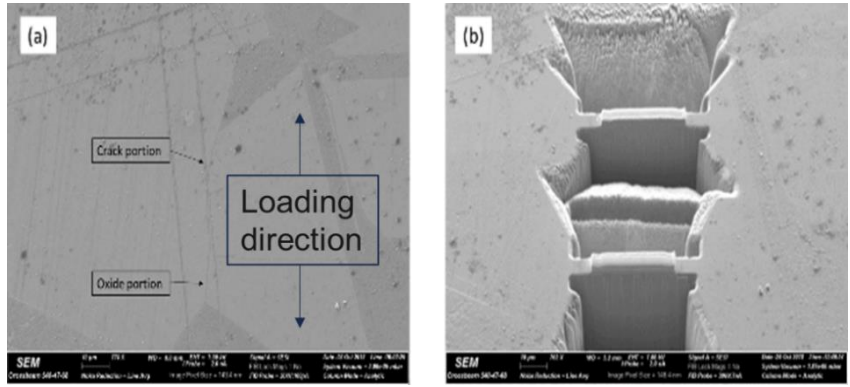

Figure 1: SEM SE images showing (a) the location of selected grain boundary with crack and oxidized portions (b) after making trench for TEM lift out.

### 2.2.2 TKD analysis

In this study we have used a new TKD configuration developed recently called "on-axis" TKD [34,35]. In this configuration there is a hardware modification compared to conventional off-axis TKD. The scintillator is moved from the vertical position in conventional TKD to a horizontal one (under the sample) in on-axis TKD, as shown in Figure 2. This enables a much faster acquisition as most of the transmitted electron are collected. It is reported that on-axis TKD can be up to 20 times faster than conventional TKD. It is also reported that on-axis TKD has much better effective lateral resolution (3-6 nm) compared to conventional TKD (10 nm) [36]. The OPTIMUS™ head also contains three forescattered electron diodes (FSD). When using the FSD imaging mode, the centre of low-angle forescattered electron diode is aligned normal to the optic axis for generating bright-field images, and the two high-angle diodes for dark-field images [37].

Kernel Average Misorientation (KAM) maps, calculated from the TKD data, show the average misorientation angle of a given point respect to all its nearest neighbors. The results are step size dependent and can sometimes be on the same order of magnitude as the measurement noise. The measurement noise is often linked to the quality of the Kikuchi patterns form which the orientation is usually determined. There are many studies in which the authors have made use of KAM to measure the plastic zones around the crack tips/ grain boundaries [18,33,38–40]. They were calculated for all the TKD datasets presented in this study.

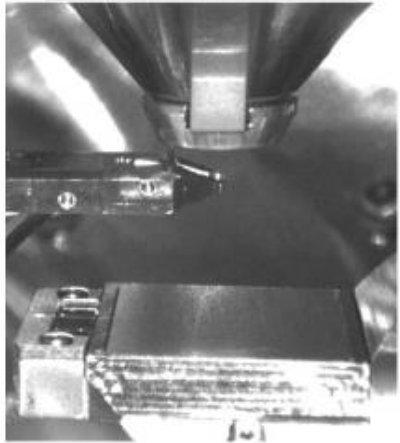

Figure 2: Chamber camera image showing the objective pole piece (top), the TKD sample holder (middle) and the "on-axis" TKD scintillator [28].

TKD maps were collected using a Zeiss Merlin FEG-SEM and eFlashHR camera with Bruker OPTIMUS$^{TM}$ detector head system. An accelerating voltage of 30 kV and a probe current of 2 nA were used. All the other microscope and acquisition parameters are listed in Table 2.

**Table 2**: Microscope and acquisition parameters

| Microscope | Zeiss Merlin FEG-SEM |
|---|---|
| Configuration | On- axis TKD |
| Camera | Bruker eFlashHR camera |
| Detector | Bruker OPTIMUS$^{TM}$ detector head |
| Accelerating Voltage | 30 kV |
| Probe current | 2 nA |
| Working distance | 5.1 mm |
| Sample tilt | 0° |
| Detector sample distance | 19.6 mm |

| Acquisition/indexation software | Bruker Esprit 2.1 |
|---|---|
| Pattern resolution | 800 x 600 pixels |
| Step size | 5 nm |
| Min. bands for indexation | 10 bands |

### 2.2.3 TKD patterns processing

Once the TKD map has been acquired, the diffraction patterns for each point on the map were saved and processed in the CrossCourt 4 software (www.hrebsd.com). Pre-processing data filters help us to obtain the KAM maps and Image Quality (IQ) maps. The analysis essentially involves picking a reference pattern (to be chosen as a point with least deformation which is away from the grain boundary) and measuring pattern shifts with respect to this point for all the other points in that grain. The patterns were then subdivided into smaller regions of interest (ROIs). In case of on-axis TKD patterns, care had been taken to avoid using ROIs close to the center of the Kikuchi pattern, where there is influence of the bright transmitted beam. The measured shifts are used to estimate elastic strain distribution and lattice rotation (which can then be used to estimate GND density content). All CrossCourt 4 GND density calculations were performed based on $L^1$ optimization method developed by Wilkinson *et al.* [23] and more details about the procedure are available in the literature [23,36].

### 2.2.4 Transmission Electron Microscopy (TEM) Characterization

TEM is the conventional tool for measuring and determining dislocation densities. TEM characterization of dislocations was performed on a JEOL 2100F $LaB_6$ microscope at 200 kV. In TEM observations some defects may be invisible at certain sample depth or diffraction conditions. To improve the visibility of the defects, we had used the convergent weak beam dark-field imaging conditions [41] , combined with electron beam precession method [42,43]. Precession of the electron beam was controlled by a script to capture 10 exposures with slightly varying deviation parameters ~0.01 nm$^1$, under weak beam conditions from 8g to 9g. The sequence of images were then aligned and stacked using ImageJ [44] to produce a single micrograph with reduced extinction distance-related absences of defects. The stacked micrographs are then shown in inverted contrast after background subtraction using the rolling-ball method (radius =40 pixels) in ImageJ [44]. More details about this technique and image acquisition procedure can be found in [45].

# 3    Results

After autoclave testing, the sample exhibited several intergranular cracks with an average depth of 5.15 ± 1 µm from the surface. These cracks are in the regime of initiation stage. Two cracked grain boundaries were selected for analysis. One of them (cracked grain boundary 1), containing a cracked and an uncracked portion. This grain boundary was at a very shallow angle (5°) respect to the loading direction, as shown in Figure 1, which probably affected its SCC mode, and where the strain analysis might provide with useful information.

### 3.1    Cracked grain boundary 1

The sample containing the cracked portion of the grain boundary was imaged using the forescatter diodes (Argus detectors) before TKD analysis (Figure 3a), providing an insight of how the dislocations are piled up along the crack flank, this was later confirmed by TEM imaging. After this initial screening, TKD maps were obtained as described previously. Figure 3b shows the Inverse Pole figure map (out of the plane) of the cracked region. The misorientation between the grains is found to be approximately 18.8°.

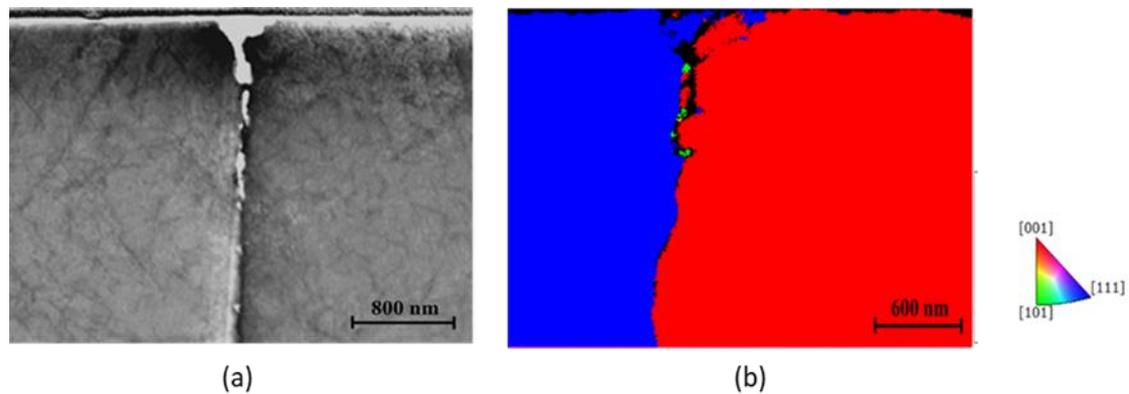

Figure 3: (a) Bright field forescattered electron image and (b) Inverse Pole Figure (IPFZ) map of cracked grain boundary 1

Figure 4 shows the elastic strain distribution around the crack region with the strain distributions in $\varepsilon_{22}$ suggesting that there is mostly tensile strain close to the crack flank. This is also in agreement with the sample lift out being in the tensile region of the 4 point bend test. With the chosen reference points, it is obvious that in the area above this point, the elastic strain distribution was tensile, while the region below was compressive. The strain paths in Figure 4 also appear to correspond to the dislocation pileups observed in the forescatter image in Figure 3(a). It will be shown later than these pileups are also oriented along the slip directions in the corresponding grain.

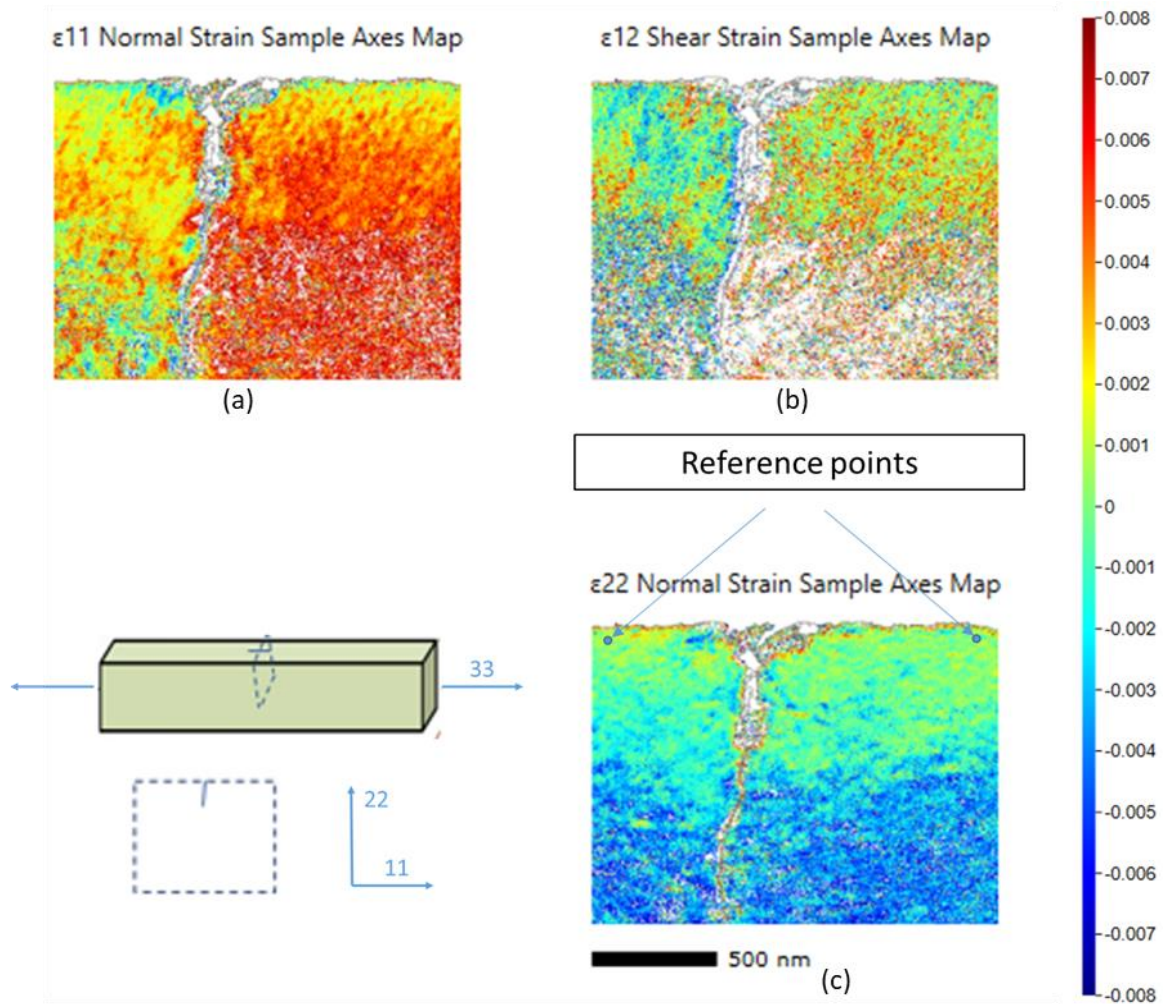

Figure 4: Elastic Strain maps around the cracked GB 1 region along (a) $\varepsilon_{11}$ (b) $\varepsilon_{12}$ (c) $\varepsilon_{22}$

### 3.2 Uncracked grain boundary 1

The sample containing the uncracked part of the grain boundary exhibited intergranular oxidation with a depth of 3 µm as seen in Figure 5 (a). It was imaged using the forescatter diodes before TKD analysis and this provided the insight of how the dislocations are piled up along the grain boundary as a result of deformation. Figure 5 shows the forescatter image and IPF-Z maps around the oxidized grain boundary region. Note that, due to detector saturation, the oxidized portion of the grain boundary appears white and similar to an open crack, but a careful examination confirmed that it was only oxidized down to 3 µm and uncracked. After this, a TKD map was obtained in a similar way as the cracked grain boundary and analyzed using Crosscourt 4 software.

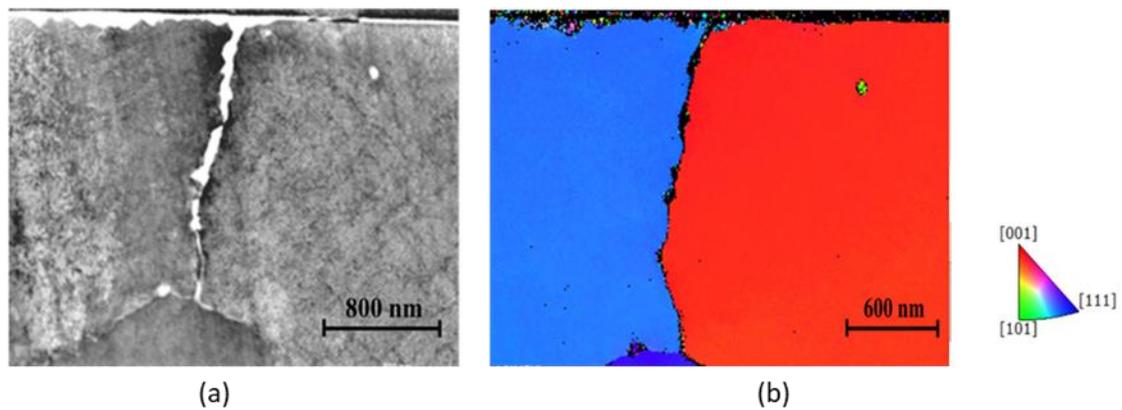

Figure 5: (a) Bright field forescattered electron image and (b) Inverse Pole Figure (IPFZ) map of the uncracked portion of GB 1

Figure 6 shows the elastic strains around the oxide grain boundary. It appears that from the $\varepsilon_{22}$ strain analysis, one grain has higher compressive strain and the other grain has higher tensile strain with reference to the grain boundary plane in the center.

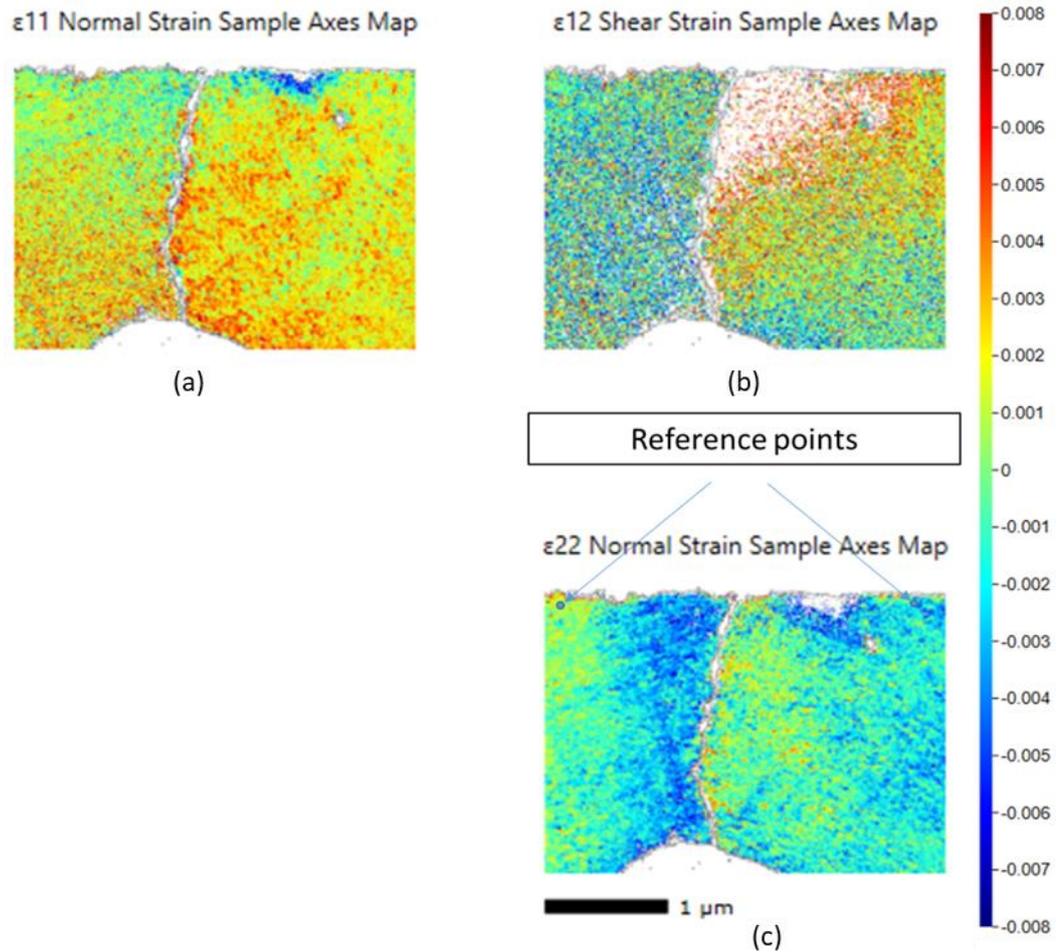

Figure 6: Elastic Strain maps around the uncracked portion of GB 1 along (a) $\varepsilon_{11}$ (b) $\varepsilon_{12}$ (c) $\varepsilon_{22}$

### 3.3 As Received grain boundary

An EBSD scan was run on the as received sample in order to find a similar grain boundary to the cracked grain boundary 1 (with a misorientation close to 18.8°). Figure 7 shows the selected grain boundary which was extracted in cross-sectional orientation as a TEM lift-out for analysis.

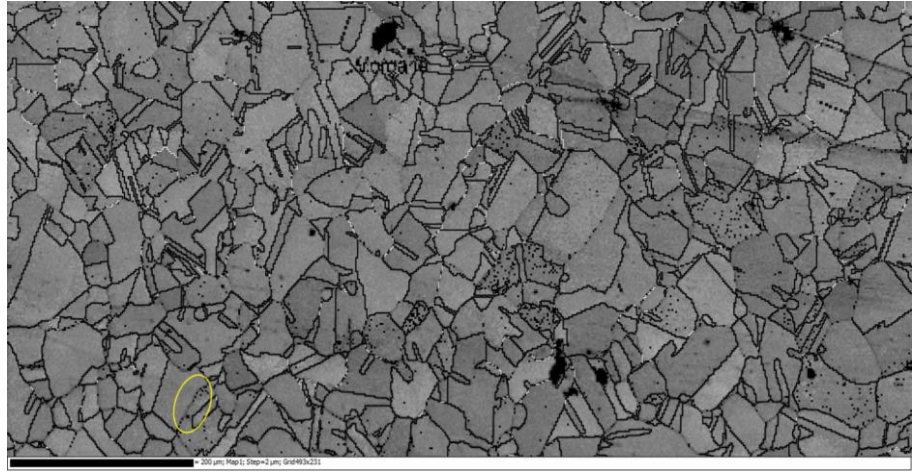

Figure 7: EBSD grain boundary map from the as-received sample showing the selected GB (circled) with a misorientation of 19°.

Using the similar routine in the two above cases, the as received sample was analyzed using Argus forescatter detector before TKD analysis. The forescatter image in Figure 8 showed no pile up of dislocations as seen in the two samples extracted from the cracked grain boundary. The elastic strain map in Figure 9 only shows regions with much lower strain, likely to be due to prior thermal treatment.

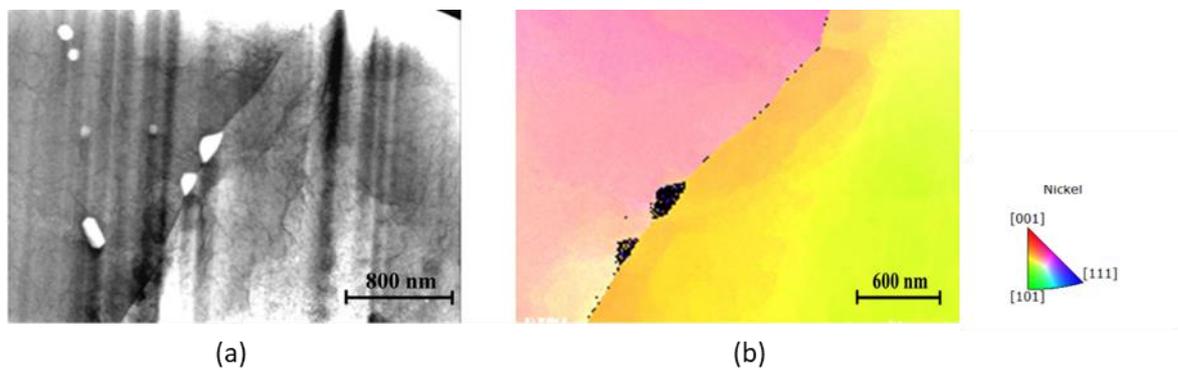

Figure 8: (a) Bright field forescattered electron image and (b) Inverse Pole Figure (IPFZ) map of as-received grain boundary region

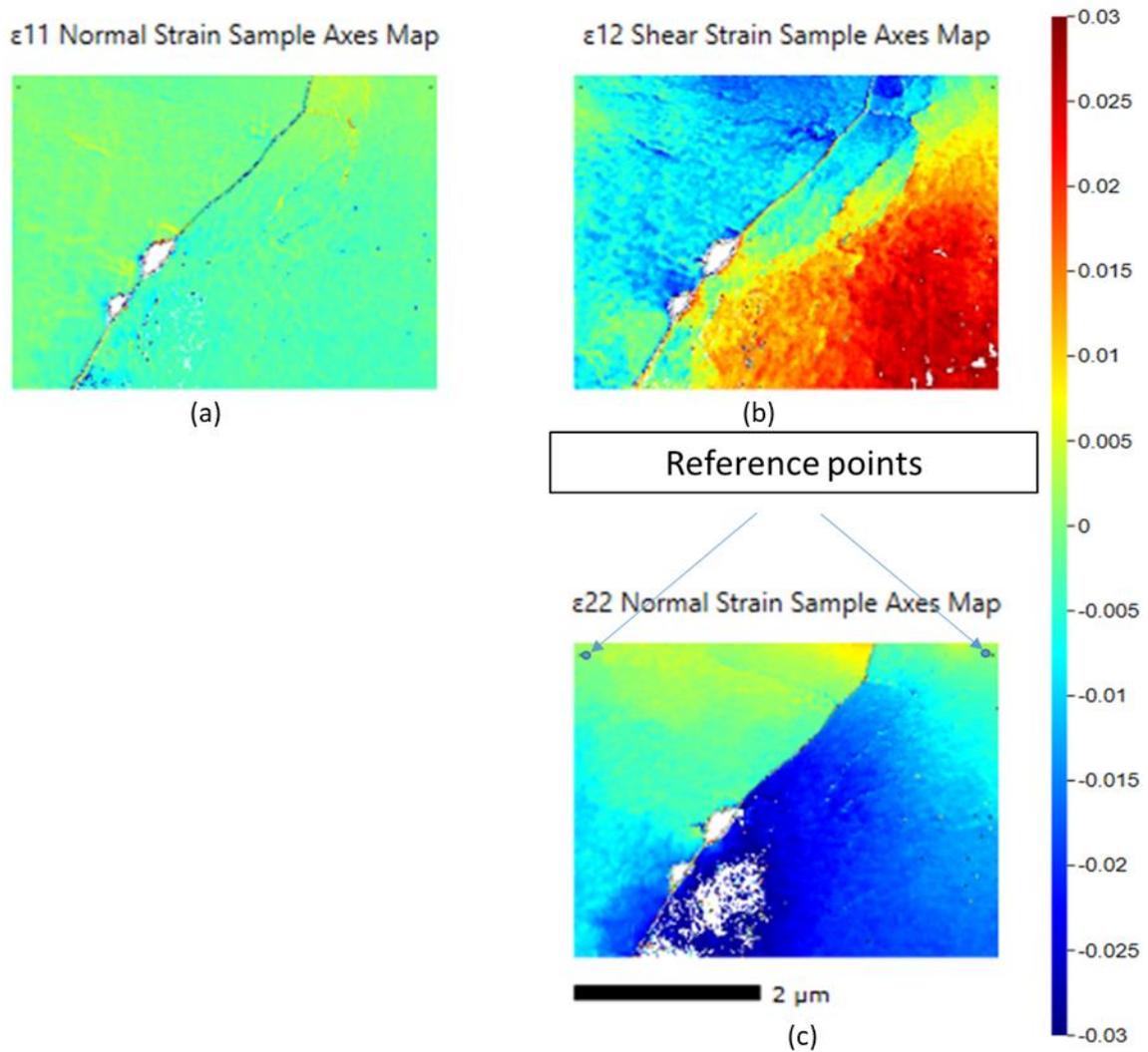

Figure 9: Elastic Strain maps around the as-received grain boundary region along (a) $\varepsilon_{11}$ (b) $\varepsilon_{12}$ (c) $\varepsilon_{22}$

### 3.4 Cracked grain boundary 2

In order to improve the significance of the measurements around cracked grain boundaries, a sample containing a different cracked grain boundary was lifted out and analyzed. Figure 10 shows the location of the crack grain boundary 2 with respect to the loading direction. The cracked grain boundary 2 was imaged using the forescatter diodes (Argus detectors) before TKD analysis (Figure 11 a), providing an insight of how the dislocations are piled up along the crack flank. After this initial screening, TKD maps were obtained as described previously. Figure 11 b shows the Inverse Pole figure map (out of the plane) of the cracked region. The misorientation between the grains was found to be approximately 17º.

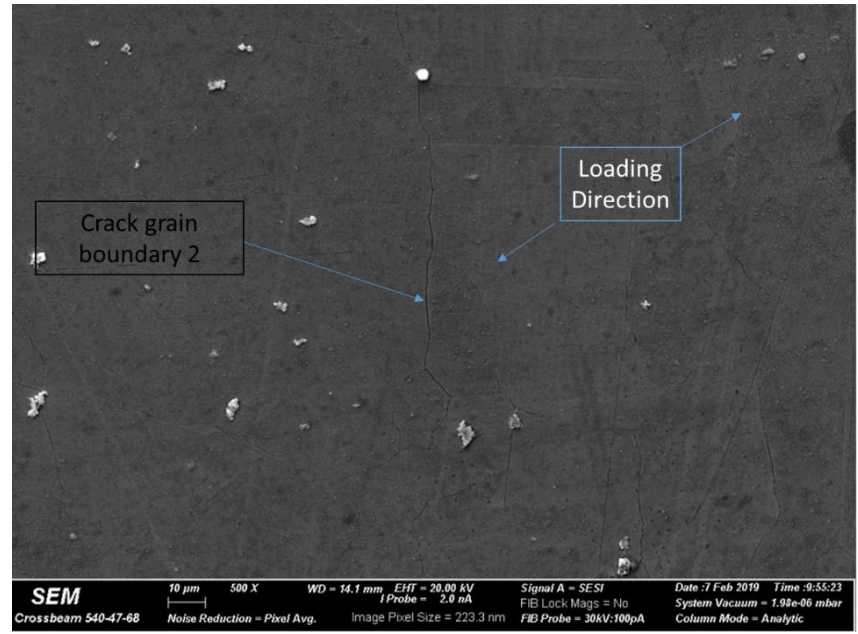

Figure 10: SEM (SE) image of the location of crack grain boundary 2 with respect to loading direction

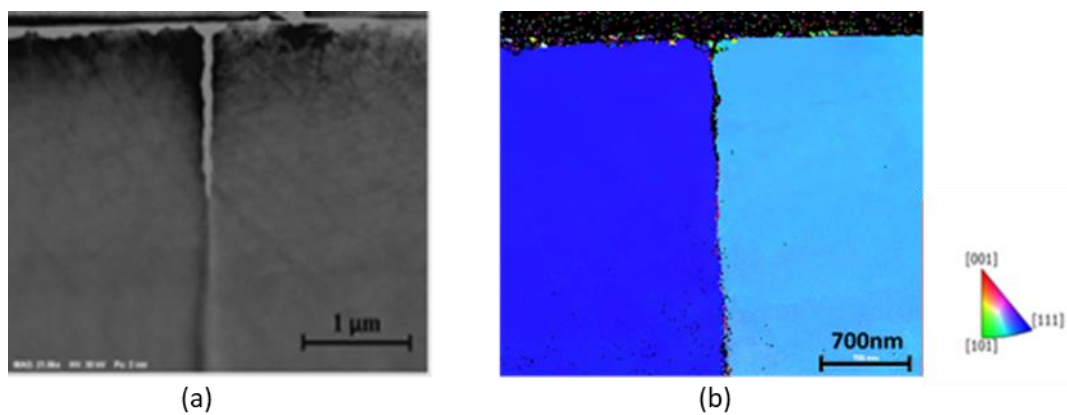

Figure 11: (a) Bright field forescattered electron image and (b) Inverse Pole Figure (IPFZ) map of crack grain boundary 2 region

Figure 12 shows the elastic strain distribution around the crack 2 region with the strain distributions in $\varepsilon_{22}$ suggesting that there was mostly tensile strain close to the crack flank. This is also in agreement with the sample lift out being in the tensile region of the 4 point bend test as seen in cracked grain boundary 1. With the chosen reference points, it is obvious that in the area above this point, the elastic strain distribution was tensile, while in the region below was compressive. The strain paths in Figure 12 also appear to correspond to the dislocation pileups observed in the forescatter image in Figure 11(a). It will be shown later than these pileups are also oriented along the slip directions in the corresponding grain.

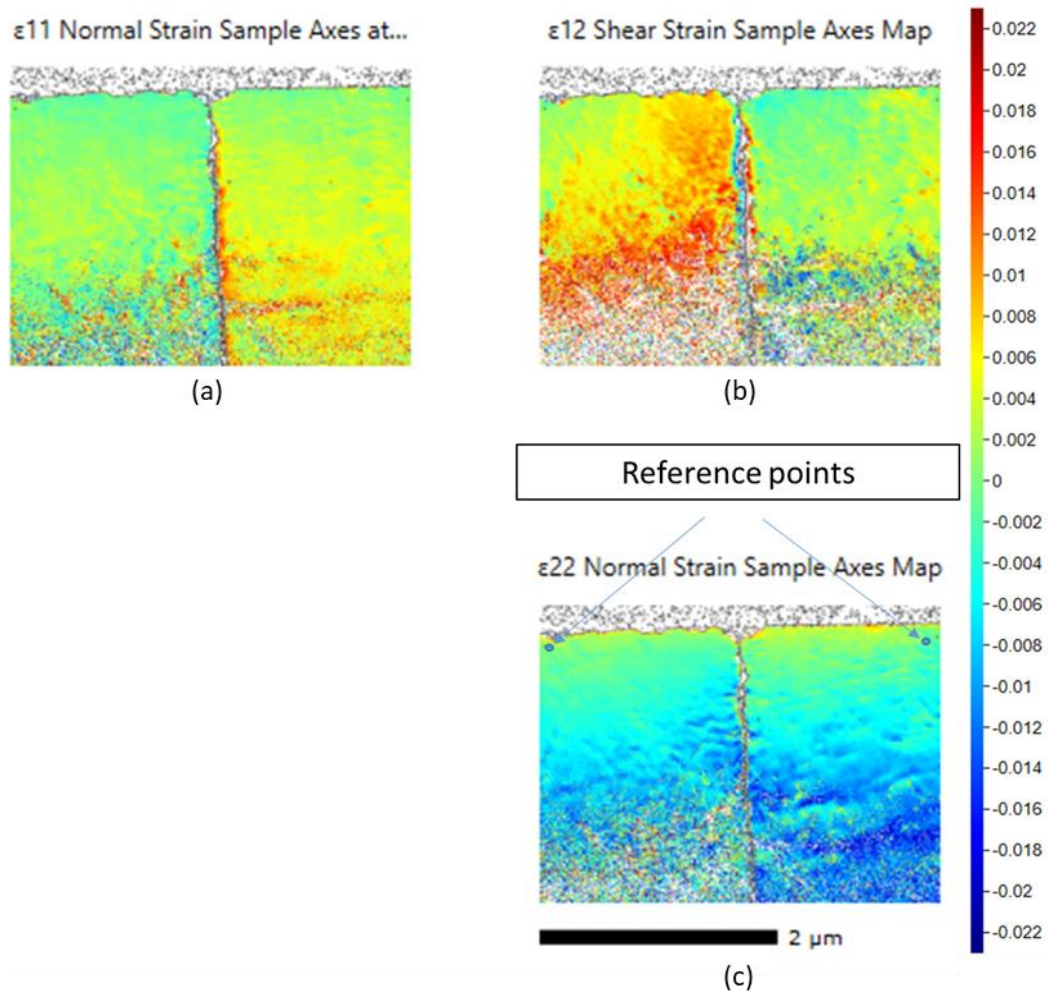

Figure 12: Elastic Strain maps around the cracked grain boundary 2 region along (a) $\varepsilon_{11}$ (b) $\varepsilon_{12}$ (c) $\varepsilon_{22}$

## 3.5 Kernel Average Misorientation (KAM) characterization

Figure 13 shows the kernel average misorientation (KAM) maps of the three samples analyzed as part of the grain boundary 1 study. Each pixel represents the misorientation average value (in degrees) with regards to all the neighboring pixels. A color temperature scale (dark blue to red) is used for easy visualization. The regions with high misorientation average appear brighter (green, yellow) than those with less misorientation average.

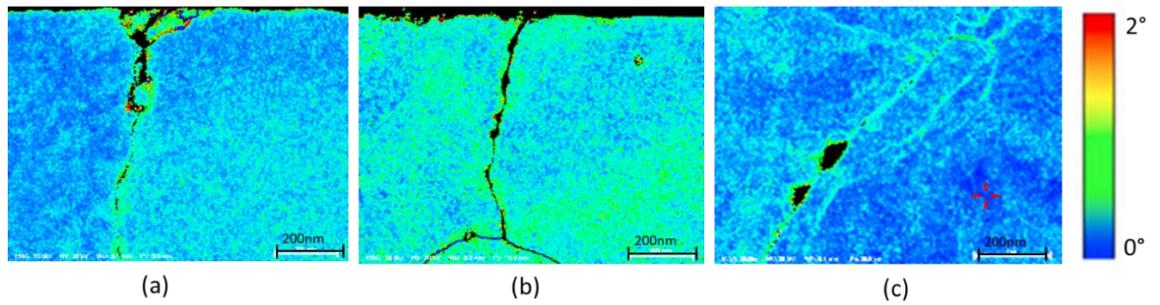

Figure 13: Kernel Average Misorientation (KAM) (a) cracked grain boundary 1 (b) oxidized grain boundary (c) as-received grain boundary

Misorientation line profiles for left and right side grain (LS and RS respectively) consistent with what reported in the literature [15,24] as shown in Figure 14 have been acquired at the crack tip region in the cracked grain boundary sample. It can be seen that only at distances of ~50 nm from the crack tip there was a misorientation gradient before a plateau was reached.

a. KAM profile of cracked GB 1

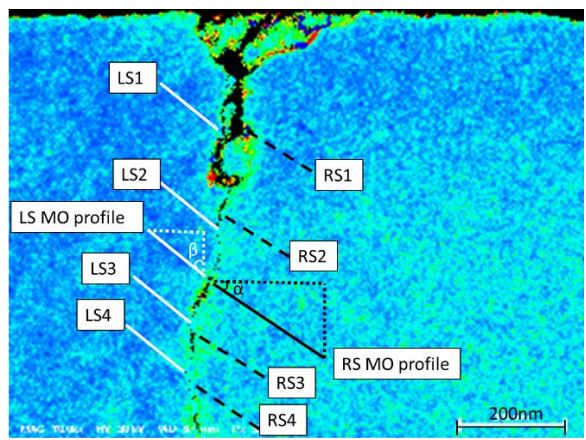

b. Left side grain (LS) profile

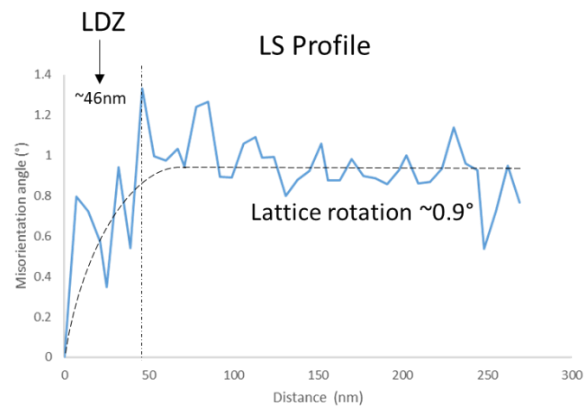

c. Right side grain (RS) profile

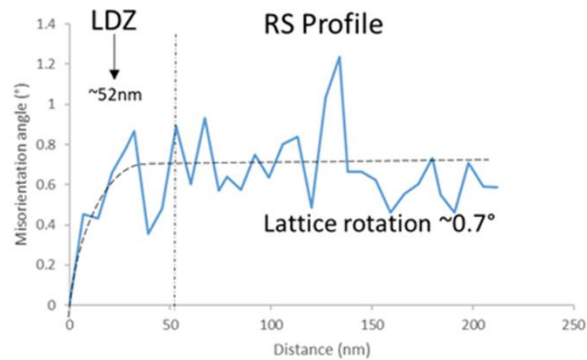

Figure 14: KAM profiles for cracked sample: (a) KAM map, location of LS profile (white line), LS1-LS4 (white dashed lines), RS profile (black line), RS1-RS4 (black dashed lines); (b) LS profile (solid blue line: raw data, dashed line: fitted data): misorientation reaches plateau at lattice rotation of 0.9°±0.3° at ~46 nm; (c) RS profile (solid blue line: raw data, dashed line: fitted data): misorientation reaches plateau at lattice rotation of 0.7° ± 0.3° at ~52 nm from the grain boundary.

The gradient in the misorientation suggests a higher lattice deformation in the region; this is called lattice deformation zone (LDZ). From this we obtain two parameters, distance from the grain boundary at which the misorientation value reaches the maximum and secondly lattice rotation which is caused by the localized deformation. In both the grains, the misorientation reached a plateau value at similar distance although to slight different rotation values. The length of the local deformation zone was 46 nm in the left grain and 52 nm in the right grain, and the plateau lattice rotation was 0.9° ± 0.3° in the left grain and 0.7° ± 0.3° in the right one. The profiles were obtained at a different angle from the grain boundary orientation (horizontal) in order to avoid the intersection with the deformation bands. Therefore, using basic trigonometry, actual comparable distances from the grain boundary, open crack and crack tip have been calculated as shown in Table 3. This results confirms a higher dislocation density localized around the crack tip/cracked grain boundary [40].

**Table 3**

Misorientation profiles: the LDZ refers to the distance from the grain boundary within which there is a misorientation gradient is present. The actual LDZ refers to the LDZ perpendicular to the grain boundary.

| Profile no. | Grain | LDZ (nm) | Angle (°) | Actual LDZ (nm) | Lattice rotation (°) |
|---|---|---|---|---|---|
| LS | Left | 46 | 50 | 35.23 | 0.9°±0.3° |
| LS1 | Left | 48 | 50 | 36.77 | 1°±0.3° |
| LS2 | Left | 30 | 50 | 22.9 | 0.6°±0.3° |
| LS3 | Left | 36 | 50 | 27.57 | 0.7°±0.3° |
| LS4 | Left | 32 | 50 | 24.51 | 0.8°±0.3° |
| RS | Right | 52 | 45 | 36.7 | 0.7°±0.3° |
| RS1 | Right | 48 | 45 | 33.94 | 1.2°±0.4° |
| RS2 | Right | 35 | 45 | 24.7 | 0.7°±0.3° |
| RS3 | Right | 40 | 45 | 28.28 | 0.6°±0.3° |
| RS4 | Right | 38 | 45 | 26.87 | 0.8°±0.3° |

**3.6    Geometrically Necessary Dislocation Density calculation**

As a result of cross correlation analysis, we also obtain lattice rotations as an output. With knowledge of dislocation types in the system, we fit the dislocation type to the measured curvatures supporting these rotations and can thus obtain an estimate of geometrically necessary dislocation density (GND density). After deformation the residual stresses remain due to the inhomogeneity of the plastic deformation throughout the grains

[40]. GNDs maps and the quantification of local dislocation densities could contribute to better understanding SCC initiation, since they could highlight some form of density threshold that would allow stress localization to reach the minimum required to fracture an oxidized grain boundary [25,26]. For this reason, after cross-correlation, dislocation density maps were calculated for all samples analyzed as part of the grain boundary 1 study. The GND density calculation were performed based on $L^1$ optimisation method developed by Wilkinson *et al.* [23] using the CrossCourt 4 software. An average value was extracted from the region close to the grain boundary (first 50 nm). The cracked grain boundary had the highest value with ~ $3 \times 10^{16}$ lines/m$^2$. The oxidized grain boundary was second with a value ~ $5 \times 10^{15}$ lines/m$^2$ whereas the as-received grain boundary had the lowest value ~$5 \times 10^{14}$ lines/m$^2$. This is expected as both should have experienced deformation during testing under 4 point bending.

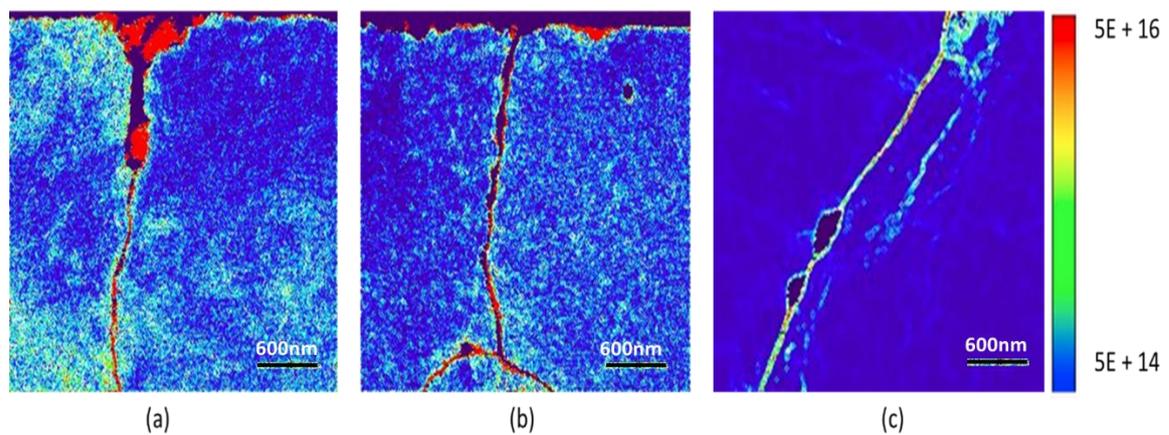

Figure 15: GNDs map (lines/m$^2$) (a) cracked grain boundary 1 (b) uncracked grain boundary 1 (c) as-received grain boundary (similar to 1).

Figure 16 shows GND density maps of the second crack. In the case of the second crack a GND density of ~$1 \times 10^{16}$ lines/m$^2$ was measured. Thus, from GNDs of both the cracked regions, the minimum observed GNDs around the grain boundaries that failed (initiating cracking) was found to be more than $5 \times 10^{15}$ lines/m$^2$.

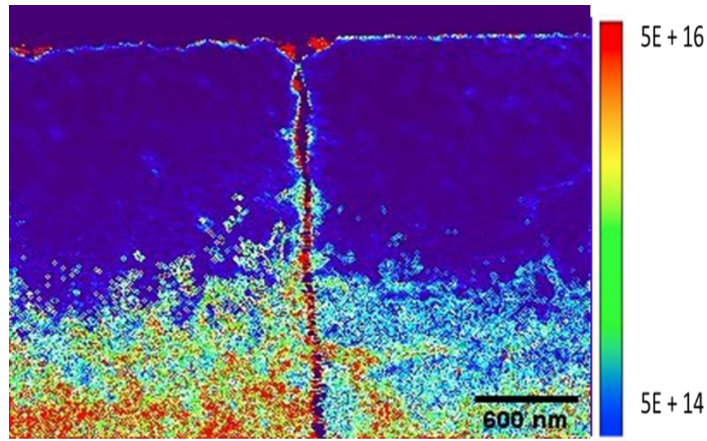

Figure 16: GND density maps of crack grain boundary 2 (lines/m$^2$).

GND line profiles similar to the KAM analysis ones (Figure 14) were extracted from the GND map of the cracked grain boundary 1, as shown in Figure 17. The line profiles start at the grain boundary (crack) and were averaged over a width of 50 nm perpendicular to the profile (see Figure 17 a), in order to reduce the influence of local noise. The GND profiles from each of the flanks of the cracked grain boundary (Figures 17 b, 17 c) show that there was a higher GND density in the first 20-40 nm after the crack before reaching a plateau; this was within the region of the LDZ. Since the lattice rotation gradient can be directly related to the dislocation densities [25], it is possible to quantitatively estimate the maximum GNDs in the LDZ because these two parameters are very much correlated.

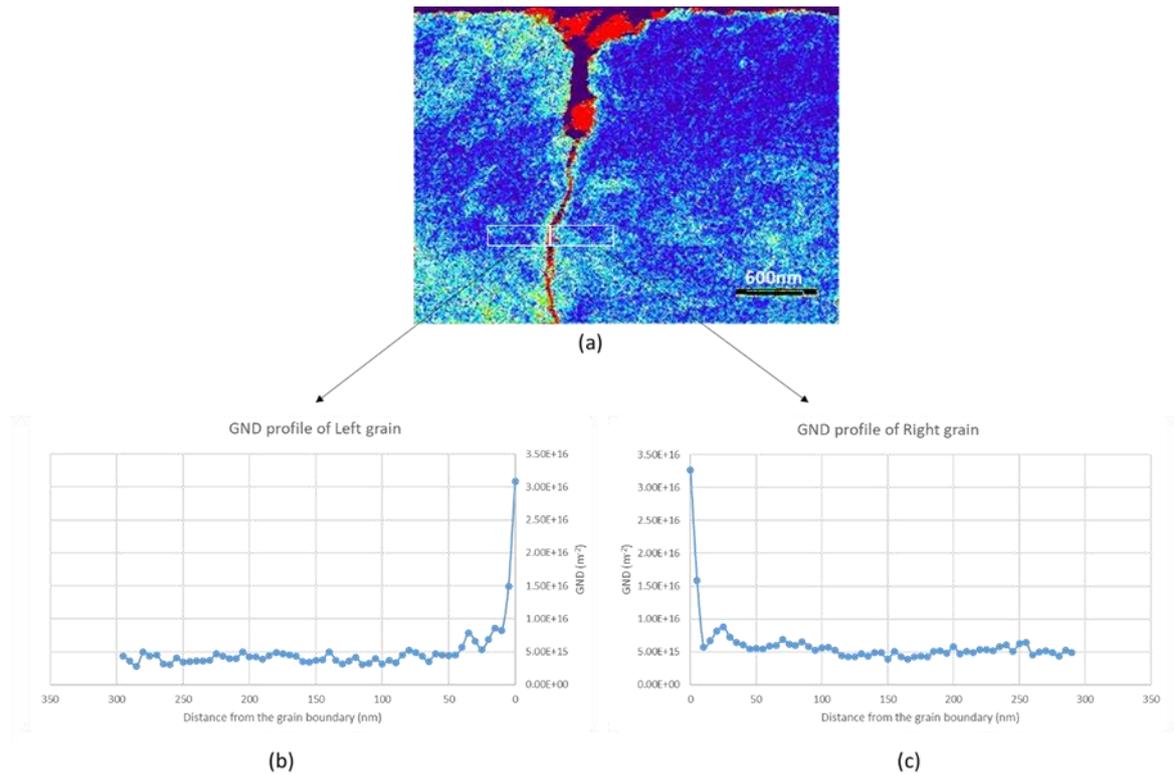

Figure 17: GND map containing the cracked grain boundary (a) GND map with indication of the line profile positions. (b) GND line profile left grain (c) GND line profile right grain

### 3.7 Dislocations and Schmidt factor calculation

Dislocation accumulation, pileups around microstructural features play a crucial role in crack initiation and propagation. Microstructural investigation at this scale gives an insight into the complex nature of the local deformation. TKD investigation at this scale offers knowledge about the dislocation and strain accumulation around the microstructural features of interest.

In the case of cracked grain boundary, there is significant amount of strain along the crack flank on the left grain. The components of the elastic strains is of order $10^{-3}$ which is relatively large. This is evident from the GND density maps as region of elevated GND content and it was proved by transmission electron microscopy (TEM) analysis of dislocations in this region. TEM analysis along this region showed in Figure 18 that the dislocations were straight and along the same plane as shown in Figure 3.

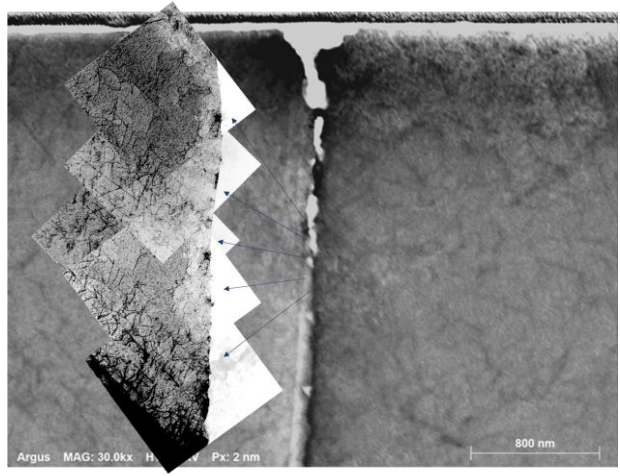

Figure 18: Montage of dislocations using TEM imaging close to the cracked grain boundary 1.

These slip pileups were seen to occur along the {111} direction in the crystal. Traces of <111> planes are shown in Figure 19 and they correspond to the pileup directions. This is in good agreement with the fcc slip system [39,46].

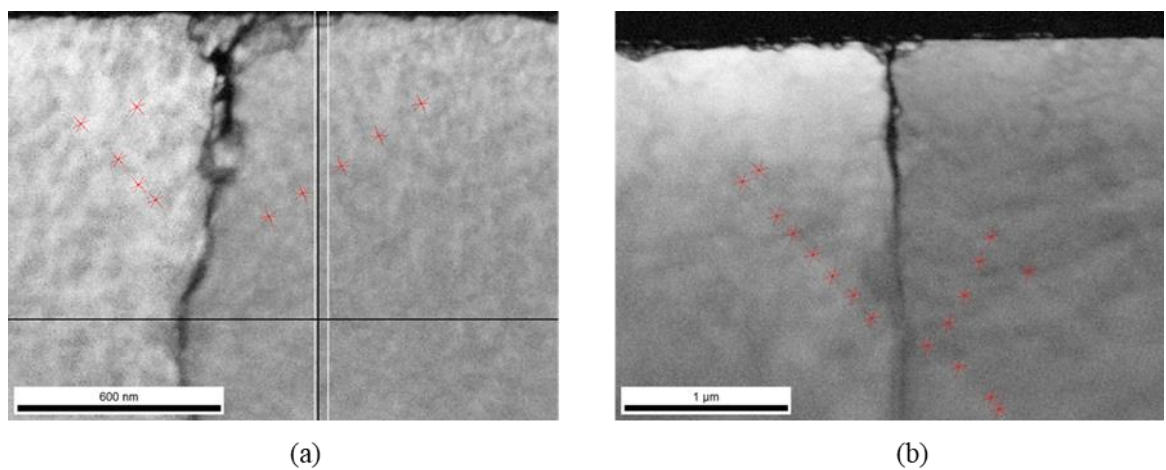

Figure 19: Slip plane tracing using TSL OIM software on Image quality map (white and black lines mark the {100} direction and <100> plane trace and the red markers point the direction of the slip planes {111} in (a) crack GB 1 & (b) crack GB 2).

Dislocations will not move along the slip plane until a critical resolved shear stress has been reached. This brings into account of Schmid factor [47]. The Schmid factor does not directly describe slip within a grain, however, it does examine the resolved shear stress acting on a single slip system, which is important when considering the critical resolved shear stress needed to initiate slip on that slip system. The Schmid factor was found to be 0.45 for the grains adjacent to the two cracked grain boundaries in this study. Clear evidence of slip lines can be observed in Figure 19, indicating that the applied stress is greater than the critical resolved shear stress value. This is similar to what previously reported in the literature for irradiation assisted SCC IASCC [48–50].

## 4. Discussion

In this study, three grain boundaries and four different regions are analyzed using a novel on-axis TKD technique. This technique provides a remarkable spatial resolution allowing to orientation mapping and lattice deformation to be analysed in a better way.

The forescatter images (Figures 3, 5, 13) are an effective way of identifying the defects in the material. The regions of higher dislocation density or deformation zone appear darker, whereas crack and oxides appear brighter due to the contrast settings. Hence, with these forescatter images it is possible to locate the SCC crack, grain boundary, deformation bands, slip planes and dislocation pile ups with higher resolution.

Figure 9 shows the strain distributions in a grain boundary similar to GB 1, in the as-received sample, before it was tested. It is worth mentioning that there already existent strains around boundaries or inclusions were non-homogenously distributed as the result of cooling after annealing. A similar situation was observed in the uncracked portion of GB 1 (Figure 6). The strain components parallel to the GB ($\varepsilon_{22}$) exhibited a gradient towards the grain boundary. Using a reference point about 1 micron from the GB, one grain appeared to be under compression and the other under tension the region close to the GB. It is difficult to see how for neighboring grains one grain is in compression and the other is in tension, unless there is a complex stress condition and constraint from the surrounding matrix. Once the grain boundary has failed (Figure 4), a strain gradient was observed in the direction parallel to the grain boundary, where the upper part of the grain (next to the free surface) appeared to have relaxed due to the crack. This behavior was observed in the crack grain boundary 2 (Figure 12) as well, where the upper part appeared to be relaxed.

The alloy was deformed plastically and unloaded after the 4 point bend test in autoclave, bringing in the geometrically necessary dislocations to maintain the lattice continuity [39][46]. After such deformation the residual stresses remained due to the inhomogeneity of the plastic deformation throughout the grains [40]. In order to have a realistic GND density reference to compare to, GND densities in the as-received sample (Figures 9 and 14 c) were measured, with an average value of $5 \times 10^{14}$ m$^{-2}$.

By studying the GBs that failed (including their uncracked portion), local GND density values could be obtained, revealing any potential threshold above which an oxidized GB would fail. Cracked GB 1 was found to have the highest local GND density, in the order of $3 \times 10^{16}$ m$^{-2}$, crack GB 2 has a value of $1 \times 10^{16}$ m$^{-2}$ and the uncracked portion of GB 1 exhibited values in the order of $5 \times 10^{15}$ m$^{-2}$. This is expected as both cracked GBs have experienced deformation during testing. The amount of GND observed in all these cases could be a combination of both the intergranular oxidation and mechanical deformation.

The crosscorrelation analysis of TKD patterns provides us the means of investigating the dislocation arrangements at a microscopic scale. At this scale many important phenomena occur during plastic deformation as a result of complex interactions between dislocations and microstructural features. In the case of the cracked grain boundaries, there is significant amount of strain gradient in the normal direction to the free surface. The correlation between local misorientation and strain carried out previously by Sasaki *et al.* [18] and Ilevbare *et al.* [19] using DIC already showed that the measured local strains (elastic and plastic) are higher around cracked GBs, although with much lower spatial resolution. In addition, DIC can only measure the strain on the surface of the sample. The SEM based DIC and other similar techniques do only ascertain in extracting or estimating the strain on the surface. Whereas, the TKD coupled with cross correlation helps in evaluation of the strain into the sample as SCC cracks propagate cross sectional as well. DIC for this study would prove to be really difficult to be implemented because the corrosion on the sample surface would destroy or modify any surface pattern whose stability is paramount for the successful implementation of digital image correlation. Since SCC initiation and propagation are local phenomena, they are likely to be influenced by changes in the microstructure or strain distribution around the crack tip and, therefore, at a much smaller scale. Using TKD we can accurately calculate the elastic strain and GND densities with nm resolution, resulting in more realistic values of strain than the previously reported [19]. GND analysis has the potential of becoming a more reliable technique than DIC and KAM analysis for SCC initiation studies, since it is sensitive to small strain changes and, more importantly, happening in volumes of just a few nanometers. The effect of sample thinning on strain and GND measurements is still under investigation, although preliminary results suggest that both measurements are still fully representative of the bulk sample [51]. These observations support the conclusion that SCC initiation is controlled by local deformation.

## 5. Conclusions

A significant amount of microstructural and micromechanical features related to SCC have been analyzed successfully using on-axis TKD, a novel high resolution orientation mapping method. Several grain boundaries have been analyzed and their cracked and uncracked portions were compared. All cracked boundaries analyzed had local GND densities higher than $1 \times 10^{16}$ m$^{-2}$. Similar grain boundaries, from as-received samples had GNDs of $5 \times 10^{14}$ m$^{-2}$, while an intermediate level was found in the oxidized but uncracked portion of the same GB. This would suggest that a threshold value for the local GND density is required to generate the stress required for an oxidized grain boundary to fail (SCC initiation).

Strain maps were used to better understand the evolution of local strain distributions around grain boundaries before and after cracking, which might prove useful when validating finite element modelling results. It was also shown that high-resolution strain and GND maps have the potential to contribute to a better understanding of crack initiation and propagation.

**Acknowledgements**

The authors would like to thank EDF R&D, Avenue des Renardieres for providing the sample and Dr. Jonathan Duff, University of Manchester for helping us with autoclave testing. The EPSRC (EP/K040375/1, EP/N010868/1 and EP/R009392/1) and EDF MAI-SN grants are also acknowledged for funding this research.